\documentclass[a4paper,12pt, epsfig]{article}
\pdfoutput=1
\usepackage{epsfig}
\usepackage{epstopdf}
\usepackage{graphicx}
\usepackage{ifthen}
\usepackage{feynmp}
\usepackage{amsmath}
\usepackage[psamsfonts]{amssymb}
\usepackage{euscript}

\usepackage{latexsym}
\usepackage[arrow,matrix,curve]{xy}

\newskip\humongous \humongous=0pt plus 1000pt minus 1000pt

\newif\ifdtup

\def\,{\hspace{-.1cm}}
\def\hsp{,\hspace{.7cm}}

\def\fc#1#2 {\frac{n}{q}#1\frac{n}{q}#2}

\renewcommand{\cos}{\textrm{cos}}
\renewcommand{\sin}{\textrm{sin}}

\renewcommand{\sinh}{\textrm{sinh}}
\renewcommand{\cosh}{\textrm{cosh}}
\renewcommand{\tanh}{\textrm{tanh}}
\newcommand{\sech}{\textrm{sech}}
\newcommand{\csch}{\textrm{csch}}

\renewcommand{\theequation}{\arabic{section}.\arabic{equation}}
\renewcommand{\(}{\begin{equation}}
\renewcommand{\)}{end{equation} \vspace{-.05in}\linebreak}

\newcounter{saveeqn}
\newcounter{savealpheqn}

\newcommand{\alpheqn}{\setcounter{saveeqn}{\value{equation}}%
  \stepcounter{saveeqn}\setcounter{equation}{0}%
  \renewcommand{\theequation}{\mbox{\arabic{section}.\arabic{saveeqn}
\alph{equation}}}
  \renewcommand{\)}{\end{equation}}}
\def\part#1{\frac{\partial}{\partial{#1}}}%
\def\group#1{\refstepcounter{equation}\setcounter{saveeqn}
 {\value{equation}}%
  \label{#1}\setcounter{equation}{0}%
\renewcommand{\theequation}{\mbox{\arabic{section}.\arabic{saveeqn}
\alph{equation}}}
  \renewcommand{\)}{\end{equation}}}
\newcommand{\reseteqn}{\setcounter{equation}{\value{saveeqn}}%
  \renewcommand{\theequation}{\arabic{section}.\arabic{equation}}%
  \renewcommand{\)}{\end{equation}}}

\newcommand{\aalpheqn}{\setcounter{saveeqn}{\value{equation}}%
  \stepcounter{saveeqn}\setcounter{equation}{0}%
  \renewcommand{\theequation}{\mbox{
        \Alph{subsection}.\arabic{saveeqn}\alph{equation}}}
   \renewcommand{\)}{\end{equation}}}
\newcommand{\areseteqn}{\setcounter{equation}{\value{saveeqn}}%
  \renewcommand{\theequation}{\Alph{subsection}.\arabic{equation}}%
  \renewcommand{\)}{\end{equation}}}

\renewcommand{\thefootnote}{\alph{footnote}}
\renewcommand{\(}{\begin{equation}}
\renewcommand{\)}{\end{equation}}
\newcommand{\ba}{\begin{eqnarray}}
\newcommand{\ea}{\end{eqnarray}}
\newcommand{\cbp}{\mathop{\vtop{\ialign{##\crcr
   $\hfil\displaystyle{}\hfil$\crcr\noalign{\kern-13pt\nointerlineskip}
   \BIG{)}\hskip0pt\crcr\noalign{\kern3pt}}}}}
\newcommand{\pa}{\mathop{\vtop{\ialign{##\crcr

$\hfil\displaystyle{\oplus}\hfil$\crcr\noalign{\kern+1pt\nointerlineskip
}
   \hspace{.08in}$^{\alpha=0}$\hskip6pt\crcr\noalign{\kern3pt}}}}}
\renewcommand{\hsp}{,\hspace{.3in}}
\newcommand{\p}{^\prime}

\newcommand{\Z}{\ensuremath{\mathbb Z}}

\catcode`\@=11
\def\vereq#1#2{\lower3pt\vbox{\baselineskip1.5pt \lineskip1.5pt
\ialign{$\m@th#1\hfill##\hfil$\crcr#2\crcr\sim\crcr}}}
\catcode`\@=12

\renewcommand{\(}{\begin{equation}}
\renewcommand{\)}{\end{equation}}

\def\pin#1{\int \frac{d#1}{2\pi}}

\def\df{\mathcal{D}_f}

\newcommand{\beas}{\begin{eqnarray*}}
\newcommand{\eeas}{\end{eqnarray*}}

\newcommand{\bquo}{\begin{quote}}
\newcommand{\enqu}{\end{quote}}

\renewcommand{\Z}{{\mathbb Z}}

\def\ch{{\mathcal{H}}}
\def\co{{\mathcal{O}}}

\newcommand{\beq}{\begin{equation}}
\newcommand{\eeq}{\end{equation}}
\newcommand{\bea}{\begin{eqnarray}}
\newcommand{\eea}{\end{eqnarray}}

\newskip\humongous \humongous=0pt plus 1000pt minus 1000pt

\newif\ifdtup

\catcode`\@=11

\@addtoreset{equation}{section}

\def\@normalsize{\@setsize\normalsize{15pt}\xiipt\@xiipt
\abovedisplayskip 14pt plus3pt minus3pt%
\belowdisplayskip \abovedisplayskip
\abovedisplayshortskip \z@ plus3pt%
\belowdisplayshortskip 7pt plus3.5pt minus0pt}

\def\small{\@setsize\small{13.6pt}\xipt\@xipt
\abovedisplayskip 13pt plus3pt minus3pt%
\belowdisplayskip \abovedisplayskip
\abovedisplayshortskip \z@ plus3pt%
\belowdisplayshortskip 7pt plus3.5pt minus0pt
\def\@listi{\parsep 4.5pt plus 2pt minus 1pt
      \itemsep \parsep
      \topsep 9pt plus 3pt minus 3pt}}

\relax

\catcode`@=12

\catcode`\@=11

\def\section{\@startsection{section}{1}{\z@}{3.5ex plus 1ex minus  .2ex}{2.3ex plus .2ex}{\large\bf}}

\def\thesection{\arabic{section}}
\def\thesubsection{\arabic{section}.\arabic{subsection}}

\def\appendix{\setcounter{section}{0}
 \def\thesection{Appendix \Alph{section}}
 \def\thesubsection{\Alph{section}.\arabic{subsection}}
 \def\theequation{\Alph{section}.\arabic{equation}}}
\renewcommand{\theequation}{\arabic{section}.\arabic{equation}}

\begin{document}
\def\thefootnote{\fnsymbol{footnote}}
\def\thetitle{Finite Derivation of the One-Loop Sine-Gordon Soliton Mass}
\def\autone{Jarah Evslin}
\def\auttwo{Hengyuan Guo}
\def\affa{Institute of Modern Physics, NanChangLu 509, Lanzhou 730000, China}
\def\affb{University of the Chinese Academy of Sciences, YuQuanLu 19A, Beijing 100049, China}

\begin{center}
{\large {\bf \thetitle}}

\bigskip

\bigskip

{\large \noindent  \auttwo{${}^{1,2}$}\footnote{guohengyuan@impcas.ac.cn} and \autone{${}^{1,2}$} \footnote{jarah@impcas.ac.cn}}

\vskip.7cm

1) \affa\\
2) \affb\\

\end{center}

\begin{abstract}
\noindent
Calculations of quantum corrections to soliton masses generally require both the vacuum sector and the soliton sector to be regularized.  The finite part of the quantum correction depends on the assumed relation between these regulators when both are taken to infinity.  Recently, in the case of quantum kinks, a manifestly finite prescription for the calculation of the quantum corrections has been proposed, which uses the kink creation operator to relate the two sectors.  In this note, we test this new prescription by calculating the one-loop correction to the Sine-Gordon soliton mass, reproducing the well-known result which has been derived using integrability.


\end{abstract}

%
\setcounter{footnote}{0}
\renewcommand{\thefootnote}{\arabic{footnote}}

\section{Introduction}
In general, quantum corrections to soliton masses can be computed using the WKB approximation introduced in Ref.~\cite{dhn2}.  In Ref.~\cite{dhnsg} this method was applied to the Sine-Gordon soliton and it was found to yield the exact answer of \cite{colemansg}, as was confirmed using integrability in Ref.~\cite{luther}.   

The soliton mass is defined to be the difference between the lowest energy configurations in the one-soliton and vacuum sectors.  These two energies are themselves both infinite, and so both must be regularized and then the regulators must be taken to infinity.  The result of this calculation depends on the relation between the regulators when this limit is taken \cite{re}, and it is in general not known which relation yields the right answer.  For example, identifying modes in a compactified theory yields a different mass than an identification of momentum cutoffs. Supersymmetric and integrable models are the exception, as the soliton mass can be computed using supersymmetry and integrability and so one can determine which relation between regulators agrees with this answer.  For example a regulator which preserves the supersymmetry is guaranteed to yield the correct answer.  Therefore it may appear as though the WKB method can only be used to compute soliton masses which are already known.

A resolution to this problem was proposed in Ref.~\cite{mekink}.  It was noted that the vacuum and one-soliton sectors are related by the operator which creates the soliton, and so this operator provides the correct identification of the regulators.  As scalar theories in 1+1 dimensions can be rendered finite by normal-ordering, the vacuum Hamiltonian was normal ordered and corresponding one-soliton sector Hamiltonian was directly computed using this identification.  The one-soliton sector Hamiltonian was not normal ordered when written in terms of the eigenfunctions of its kinetic term, but simply commuting the corresponding creation operators to the left produced a constant term which was precisely equal to the result of Ref.~\cite{dhn2} for the one-loop correction to the mass.

In this paper we test the method introduced in Ref.~\cite{mekink} to derive the one-loop correction to the mass of the Sine-Gordon soliton.  This correction has been derived using integrability in Ref.~\cite{luther}, with no arbitrary choice of regulator matching, and so it provides a robust test of the method. 

First of all, we shift the scalar field by the classical soliton solution to derive the one-soliton sector Hamiltonian.   We find that only the quadratic terms contribute to the soliton mass at one-loop and we identify these terms with the Poschl-Teller Hamiltonian.  We use the classical solutions of this Hamiltonian to exactly diagonalize it, providing the desired soliton mass as well as the Hamiltonian describing the excited states in the soliton sector as a sum of quantum harmonic oscillator states.


\section{ P\"oschl-Teller Potential} \label{ptsez}

\subsection{Vacuum State and the Soliton}

The Sine-Gordon Hamiltonian is
\beq
H=\int dx \ch(x) \hsp
\ch(x)=\frac{1}{2}:\pi(x)\pi(x):+\frac{1}{2}:\partial_x\phi(x)\partial_x\phi(x):-\frac{m^2}{\lambda}:\left(\cos(\sqrt{\lambda}\phi(x))-1\right):\label{hsq}
\eeq
where $m$ and $\lambda$ are positive numbers.  The field $\phi$ has dimensions of [action]${}^{1/2}$, $m$ has dimensions of [mass] and $\lambda$ has dimensions of [action]${}^{-1}$ therefore the only dimensionless constant is $\lambda\hbar$.  Our loop expansion will therefore be an expansion in $\lambda\hbar$.  We however set $\hbar=1$ everywhere.  

The theory has a series of  degenerate ground states $|0\rangle_k$ with
\beq
{}_k\langle 0|\phi|0\rangle_k=\frac{2\pi}{\sqrt{\lambda}}k\hsp k\in\Z
\eeq
and without loss of generality we will be interested in solitons which connect the adjacent ground states $|0\rangle_0\rm{\ and\ }|0\rangle_1$.



Performing the standard expansion about the ground state $|0\rangle_0$
\beq
\phi(x)=\pin{p}\frac{1}{\sqrt{2\omega_p}}\left(a^\dag_p+a_{-p}\right)e^{-ipx}\hsp
\pi(x)=i\pin{p}\frac{\sqrt{\omega_p}}{\sqrt{2}}\left(a^\dag_p-a_{-p}\right)e^{-ipx} \label{osc}
\eeq
where
\beq
\omega_p=\sqrt{m^2+p^2}
\eeq
the canonical commutation relations satisfied by $\phi$ and $\pi$ imply
\beq
[a_p,a^\dag_q]=2\pi\delta(p-q).
\eeq
The normal ordering in Eq.~(\ref{hsq}) is defined with respect to this $a$ and $a^\dag$.

Let $E_0$ and $E_K$ be the Hamiltonian eigenvalues of the vacua $|0\rangle_k$ and the one-soliton sector ground state $|K\rangle$ 
\beq
H|0\rangle_k=E_0|0\rangle_k\hsp
H|K\rangle=E_K|K\rangle. \label{scheq}
\eeq
The soliton mass is defined to be
\beq
M_K=E_K-E_0.\label{a}
\eeq
$E_0$ can be calculated in perturbation theory as in Ref.~\cite{hui}.  The leading contributions appear at two loops and are of order $O(\lambda^2)$.  We will see that they are therefore not relevant to the one-loop soliton mass which is of order $O(\lambda^0)$.  Therefore, at the one-loop order considered here, $E_0=0$.  

The classical equation of motion derived from (\ref{hsq}) is
\beq
\frac{\partial^2\phi_{cl}(x,t)}{\partial t^2}-\frac{\partial^2\phi_{cl}(x,t)}{\partial x^2}=-\frac{m^2}{
\sqrt{\lambda}}\sin\left(\sqrt{\lambda}\phi_{cl}(x,t)\right)
\eeq
which has a stationary soliton solution
\beq
\phi_{cl}(x,t)=f(x)\hsp
f(x)=\frac{4}{\sqrt{\lambda}}\arctan{e^{mx}}. \label{ksol}
\eeq
At leading order in the semiclassical expansion one expects that this will be the form factor of the soliton ground state \cite{taylor78}
\beq
\langle K|\phi(x)|K\rangle=f(x)+O(\hbar).  \label{ff}
\eeq

\subsection{Shifted Hamiltonian }

Following Ref.~\cite{hepp}, Eq.~(\ref{ff}) would be solved if $|K\rangle=\df|0\rangle_0+O(\hbar)$  where $\df$ is the displacement operator
\beq
\df={\rm{exp}}\left(-i\int dx f(x)\pi(x)\right) \label{df}
\eeq
which satisfies \cite{mekink}
\beq
[\df,\phi(y)]=-f(y)\df\hsp
:F\left[\pi(x),\phi(x)\right]:\df=\df:F\left[\pi(x),\phi(x)+f(x)\right]: \label{fident}
\eeq
where $F$ is any function of two variables.

Eq.~(\ref{ff}) leads us to rewrite the soliton ground state as
\beq
|K\rangle=\df \co|0\rangle_0
\eeq
where $\co$ is equal to the identity plus corrections of order $O(\hbar)$.   We now define the soliton sector Hamiltonian $H_K$ by the similarity transform
\beq
H\df=\df H_K.
\eeq
Then a quick calculation shows
\beq
H_K\co|0\rangle_0
=\df^{-1}H|K\rangle_0 =E_K\co|0\rangle_0. \label{quick}
\eeq
Therefore instead of searching for the eigenstate $|K\rangle$ of $H$, we may equivalently search for the eigenstate $\co|0\rangle_0$ of $H_K$.   Although $H$ and $H_K$ are related by a similarly transformation, the second problem can be treated in ordinary perturbation theory as $\co$ is equal to the identity plus loop corrections.

$H_K$ can be evaluated using (\ref{fident})
\beq
H_K[\pi(x),\phi(x)]=H[\pi(x),\phi(x)+f(x)]
\eeq
and so
\beq
H_K=E_{cl}+\int dx \left[\ch_{PT}+\ch_I\right] \label{hdf}
\eeq
where the classical energy is
\beq
E_{cl}=\int dx\left[\frac{1}{2}\left(\partial_x f(x)\right)^2+ \frac{m^2}{\lambda}\left(1-\cos(\sqrt{\lambda}f(x))\right)\right]=\frac{8m}{\lambda} \label{ecl}
\eeq
the interaction terms are
\beq
\ch_I=\frac{m^2}{\sqrt{\lambda}}\sin(\sqrt{\lambda}f(x)) \sum_{n=1}^{\infty}\frac{(-\lambda)^n}{(2n+1)!} :\phi^{2n+1}(x):-\frac{m^2}{\lambda}\cos(\sqrt{\lambda}f(x))\sum_{n=2}^{\infty}\frac{(-\lambda)^n}{2n!} :\phi^{2n}(x):
\eeq
and the Poschl-Teller (PT) Hamiltonian density is
\beq
\ch_{PT}= \frac{:\pi^2(x):}{2}+\frac{:\partial_x\phi(x)\partial_x\phi(x):}{2}+\left(\frac{m^2}{2}-m^2{\rm{sech}}^2\left(mx\right)\right):\phi^2(x):.
 \label{hpt}
\eeq

Recall that our loop expansion is an expansion in $\lambda$.  The classical energy is of order $O(\lambda^{-1})$.  Therefore the one-loop correction will be $\lambda$-independent.  As the PT terms are $\lambda$-independent, any correction derived from them will appear at one loop.  The $\ch_I$ terms on the other hand are all of at least order $O(\lambda^{1/2})$, and so only contribute at two loops and beyond.  Thus, to calculate the one-loop soliton mass, we may drop $\ch_I$ leaving
\beq
H^\prime=E_{cl}+H_{PT}\hsp H_{PT}=\int dx \ch_{PT}. \label{clpt}
\eeq
In the remainder of this note we will explicitly diagonalize $H^\prime$ and so obtain the one-loop soliton mass as well as its excitation spectrum at one loop.

\section{Solutions to the P\"oschl-Teller Hamiltonian} \label{solsez}

In this section we will calculate the inverse Fourier transforms of the eigenfunctions of the P\"oschl-Teller wave equation.  To find the  eigenstates of $H_{PT}$, we insert the factorization Ansatz
\beq
\phi_{cl}(x,t)=\psi_k(x) e^{-i \omega_k t}
\eeq
into the corresponding classical equations of motion to obtain
\beq
0=\partial^2_x \psi_k(x)+(k^2+2m^2{\rm{sech}}^2(m x))\psi_k(x)\hsp
k^2=\omega_k^2-m^2. \label{fkeq}
\eeq
There will be a bound solution $\psi_B$ corresponding to the Goldstone mode of the soliton and also, at each $k$ an even an odd continuum solution given by the hypergeometric functions \cite{flugge}
\bea
\psi^e_k(x)&=&\cosh^{2}(m x) F\left(\frac{2+ik/m}{2},\frac{2-ik/m}{2};\frac{1}{2};-\sinh^2(m x)\right) \label{gensol}\\
\psi^o_k(x)&=&\cosh^{2}(m x)\sinh(m x) F\left(\frac{3+ik/m}{2},\frac{3-ik/m}{2};\frac{3}{2};-\sinh^2(m x)\right).\nonumber
\eea
These hypergeometric fuctions may be calculated as in the Appendix of Ref.~\cite{mekink} to obtain
\bea
F\left(\frac{2+i k}{2},\frac{2-i k}{2};\frac{1}{2};-\sinh^2(x)\right)&=&\frac{\cos(k x)-\frac{m}{k}\sin(k x)\tanh(m x)}{\cosh^2(m x)}\\
F\left(\frac{3+i k/m}{2},\frac{3-i k/m}{2};\frac{3}{2};-\sinh^2(m x)\right)&=&\frac{\left(\frac{\cos(k x)}{\cosh(m x)}+\frac{k}{m}\frac{\sin(k x)}{\sinh(m x)}\right)}{
\cosh^2(m x)(1+k^2/m^2)}.\nonumber
\eea
Substituting these back into Eq.~(\ref{gensol}) and changing the normalization by a $k$-dependent factor one obtains the solutions
\bea
\psi^e_k(x)&=&\cos(k x)-\frac{m}{k}\tanh(m x)\sin(k x)\label{psi2}\\
\psi^o_k(x)&=&\sin(kx)+\frac{m}{k}\tanh(m x)\cos(k x)\nonumber
\eea
which are normalized so that
\beq
\int dx \psi^i_{k_1} (x) \psi^j_{k_2}(x)=\pi \delta^{ij} C^2_{k_1}\delta(k_1-k_2)\hsp
C_k=\sqrt{1+m^2/k^2}\hsp i,j\in\{e,o\} \label{normpsi}
\eeq
and are real for $k$ real or imaginary.

The inverse Fourier transform of
\beq
g_k(x)=\psi^e_k(x)-i\psi^o_k(x)
\eeq
is 
\beq
\tilde{g}_k(p)=\int dx g_k(x) e^{ipx}=2\pi\delta(p-k)+\frac{\pi}{k}\csch\left(\frac{\pi (p-k)}{2m}\right) \label{gtk}
\eeq
which is normalized so that
\beq
\pin{p} {\tilde{g}}_{k_1} (p) {\tilde{g}}_{k_2}(p)=\int dx g_{k_1} (x) g_{k_2}(-x)=2\pi C^2_{k_1}\delta(k_1-k_2). \label{normp}
\eeq
The delta function results from the fact that asymptotically the eigenfunctions of $H_{PT}$ and $H_0$ (defined in (\ref{h0})) are equal.  There is no $\delta(p+k)$ term because with the coefficient in (\ref{hpt}) the PT potential is reflectionless \cite{flugge}.  

Inserting
\beq
\omega_{B}=0\hsp k_{B}=im
\eeq
into  (\ref{psi2}) one finds the bound solution
\beq
g_{B}(x)=\sech\left(mx\right)
\eeq
which corresponds to the Goldstone mode of the soliton.  It satisfies the normalization condition
\beq
\int dx |g_{B}(x)|^2=C_{B}^2\hsp C_{B}=\sqrt{\frac{2}{m}}
\eeq
and has inverse Fourier transform
\beq
\tilde{g}_{B}(p)=\int dx g_{B}(x) e^{ipx}=\frac{\pi}{m}\sech\left(\frac{\pi p}{2m}\right).  \label{gtbe}
\eeq

\section{Mode Expansion } \label{diagsez}

\subsection{PT Annihilation and Creation Operators}

To diagonlize $H_{PT}$, first we decompose it
\beq
H_{PT}=H_0+\tilde{H}_{PT}
\eeq
where $H_0$ is the usual free Hamiltonian
\beq
H_0=\int dx \left[\frac{1}{2}:\pi(x)\pi(x):+\frac{1}{2}:\partial_x\phi(x)\partial_x\phi(x):+\frac{m^2}{2}:\phi^2(x):\right]=\pin{p}\omega_p a^\dag_p a_p. \label{h0}
\eeq
Recall that the operators $a$ and $a^\dag$ were defined in (\ref{osc}) by decomposing $\phi$ and $\pi$ into plane waves, which are solutions of the wave equation corresponding to $H_0$.  To diagonalize $H_{PT}$, we instead decompose $\phi$ and $\pi$ into the basis of constant frequency solutions of the PT equation.  In particular they will contain continuum and bound state contributions
\beq
\phi(x)=\phi_C(x)+\phi_{B}(x)\hsp
\pi(x)=\pi_C(x)+\pi_{B}(x)
\eeq
which, following~\cite{mekink}, may be decomposed into the PT oscillator basis
\bea
\phi_C(x)&=&\pin{k}\frac{1}{\sqrt{2\omega_k}}\left(b_k^\dag+b_{-k}\right)\frac{g_k(x)}{C_k}\hsp \phi_{B}(x)=\phi_0 \frac{g_{B}(x)}{C_{B}}. \nonumber\\
\pi_C(x)&=&i \pin{k}\sqrt{\frac{\omega_k}{2}}\left(b_k^\dag - b_{-k}\right)\frac{g_k(x)}{C_k}\hsp \pi_{B}(x)=\pi_0 \frac{g_{B}(x)}{C_{B}} \label{pib}
\eea
where we have introduced the operators $\phi_{0}$  for $\pi_0$ which are just the position and momentum operators of the soliton.

These definitions are easily inverted
\beq 
b^\dag_k=\int dx \left[ \sqrt{\frac{\omega_k}{2}}\phi(x)-\frac{i}{\sqrt{2\omega_k}}\pi(x)\right]\frac{g^*_k(x)}{C_k}\hsp
b_{-k}=\int dx \left[ \sqrt{\frac{\omega_k}{2}}\phi(x)+\frac{i}{\sqrt{2\omega_k}}\pi(x)\right]\frac{g^*_k(x)}{C_k}
\eeq
from which one sees that the continuum $b$ operators satisfy the Heisenberg algebra
\beq
[b_{k_1},b^\dag_{k_2}]=2\pi\delta(k_1-k_2) \label{balg}
\eeq
while the bound state
\beq
\phi_0=\int dx \phi(x)\frac{g^*_{B}(x)}{C_{B}}\hsp
\pi_0=\int dx \pi(x)\frac{g^*_{B}(x)}{C_{B}}. \label{pi0int}
\eeq
satisfies the canonical algebra
\beq
[\phi_0,\pi_0]=i.
\eeq

We cannot directly write $H_{PT}$ in terms of $b$ and $b^\dag$ because it is the $a$ and $a^\dag$ operators which are normal ordered.  Thus we must first write it in terms of $a$ and then convert these to $b$.  To do this one first inverts (\ref{osc})
\beq
a^\dag_p=\int dx \left[ \sqrt{\frac{\omega_p}{2}}\phi(x)-\frac{i}{\sqrt{2\omega_p}}\pi(x)\right]e^{ipx}\hsp
a_{-p}=\int dx \left[ \sqrt{\frac{\omega_p}{2}}\phi(x)+\frac{i}{\sqrt{2\omega_p}}\pi(x)\right]e^{ipx} \label{phia}
\eeq
and decomposes the $a$ operators as
\beq
a^\dag_p=a^\dag_{C,p}+a^\dag_{BE,p}\hsp
a_p=a_{C,p}+a_{BE,p}.
\eeq
As we know $a$ as a function of $\phi$, which is a known function of $b$, we can write the Bogoliubov transformation which relates the $a$ and $b$ oscillator modes
\bea
a^\dag_{C,p}&=&\pin{k}\frac{\tilde{g}_k(p)}{2C_k}\left(\frac{\omega_p+\omega_k}{\sqrt{\omega_p\omega_k}}b_k^\dag+\frac{\omega_p-\omega_k}{\sqrt{\omega_p\omega_k}}b_{-k}\right) \label{bog}\\
a_{C,-p}&=&\pin{k}\frac{\tilde{g}_k(p)}{2C_k}\left(\frac{\omega_p-\omega_k}{\sqrt{\omega_p\omega_k}}b_k^\dag+\frac{\omega_p+\omega_k}{\sqrt{\omega_p\omega_k}}b_{-k}\right)\nonumber\\
a^\dag_{BE,p}&=&\frac{\tilde{g}_{B}(p)}{C_{B}}\left[ \sqrt{\frac{\omega_p}{2}}\phi_0-\frac{i}{\sqrt{2\omega_p}}\pi_0\right]\hsp
a_{BE,-p}=\frac{\tilde{g}_{B}(p)}{C_{B}}\left[ \sqrt{\frac{\omega_p}{2}}\phi_0+\frac{i}{\sqrt{2\omega_p}}\pi_0\right].\nonumber
\eea
Note that the delta function terms in (\ref{gtk}) can be directly integrated, using the delta function, and one sees that they do not mix $a$ with $b^\dag$.  This will imply that they do not affect the one-loop mass corrections of the soliton.

\subsection{Contributions of Continuum and Bound States}

Now we are ready to diagonalize $H_{PT}$ one term at a time.  The calculation is very similar to that in Ref.~\cite{mekink}, except that here there is no odd bound state.  Let us first decompose $H_0$ and $\tilde{H}_{PT}$ into continuum and bound state contributions
\beq
H_0=H_{C,0}+H_{B,0}\hsp \tilde{H}_{PT}=\tilde{H}_{C}+\tilde{H}_{B}.
\eeq
The continuum contribution is
\bea
H_{C,0}&=&\pin{p} \omega_p a^\dag_{C,p} a_{C,p}\nonumber\\
&=&\frac{1}{4}\pin{k}\frac{I_5(k)}{C_k^2\omega_k}+\frac{m^2}{2}\int dx\pin{k_1}\pin{k_2}\sech^2(m x)\frac{g_{k_1}(x)g_{k_2}(x)}{C_{k_1}C_{k_2}\sqrt{\omega_{k_1}\omega_{k_2}}}(b^\dag_{k_1}b^\dag_{k_2}+b_{-k_1}b_{-k_2})\nonumber\\
&&+\pin{k}\omega_k b^\dag_k b_k+m^2\int dx\pin{k_1}\pin{k_2}\sech^2(m x)\frac{g_{k_1}(x)g_{k_2}(x)}{C_{k_1}C_{k_2}\sqrt{\omega_{k_1}\omega_{k_2}}}b^\dag_{k_1} b_{-k_2} \label{hco}
\eea
where
\beq
I_5(k)=\pin{p}(\omega_p-\omega_k)^2\tilde{g}_k(p)\tilde{g}_{k}(p).
\eeq

Similarly the continuum contribution to the PT potential term is
\bea
\tilde{H}_{C}&=&-m^2\int dx\ {\rm{sech}}^2\left(m x\right) :\phi^2_C(x):\\
&=&-\frac{m^2}{8}\int dx \pin{p}\pin{q} \frac{\sech^2(\beta x)}{\omega_p\omega_q}e^{-i(p+q)x}\pin{k_1}\pin{k_2}\frac{\tilde{g}_{k_1}(p)\tilde{g}_{k_2}(q)}{C_{k_1}C_{k_2}\sqrt{\omega_{k_1}\omega_{k_2}}}\nonumber\\
&\times&\left[4\omega_p\omega_q(b^\dag_{k_1}b^\dag_{k_2}+b_{-k_1}b_{-k_2})+2\omega_q(2\omega_p+\omega_{k_1}+\omega_{k_2})b^\dag_{k_1}b_{-k_2}+2\omega_q(2\omega_p-\omega_{k_1}-\omega_{k_2})b_{-k_2}b^\dag_{k_1}\right.].\nonumber
\eea
Combining the two continuum contributions and moving all $b^\dag$ to the left using (\ref{balg}) we obtain
\beq
H_C=H_{C,0}+\tilde{H}_C=\pin{k}\omega_k b^\dag_k b_k+Q_C
\eeq
where
\bea
Q_C&=&\frac{1}{4}\pin{k}\frac{I_5(k)}{C_k^2\omega_k}+\frac{m^2}{2}\int dx\pin{p}\pin{q} \frac{\sech^2(m x)}{\omega_p}e^{-i(p+q)x}\pin{k}\frac{\tilde{g}_{k}(p)\tilde{g}_{-k}(q)}{C_{k}^2}\nonumber\\
&&-\frac{m^2}{2}\int dx\ \sech^2(m x) \pin{k}\frac{{g}_{k}(x)g^*_k(x)}{C_{k}^2\omega_{k}}.
\eea
$Q_C$ may be simplified using the equation of motion satisfied (\ref{fkeq}) by $\phi_k$ to obtain
\beq
Q_C=-\frac{1}{4}\pin{k}\pin{p}\frac{(\omega_p-\omega_k)^2}{\omega_p}\frac{\tilde{g}^2_{k}(p)}{C_{k}^2} . \label{qc}
\eeq

A similar calculation for the bound state contribution yields
\beq
H_{B}=H_{B,0}+\tilde{H}_0=\frac{\pi_0^2}{2}+Q_{B}
\eeq
where
\beq
Q_{B}
=-\frac{1}{4}\pin{p}\frac{\tilde{g}_{B}(p)\tilde{g}_{B}(p)}{C_{B}^2}\omega_p.\label{qbe}
\eeq
Using the fact that the frequency $\omega_{B}=0$ for the Goldstone mode, one sees that this is of the same form as $Q_C$ in (\ref{qc}).

\subsection{Diagonalized Hamiltonian}

Putting everything together, we have diagonalized our one-loop Hamiltonian
\beq
H_{PT}=\pin{k}\omega_k b^\dag_k b_k+
\frac{\pi_0^2}{2}+Q \label{hfin}
\eeq
where
\bea
Q&=&Q_C+Q_{B} \label{q}\\
&=&
-\frac{1}{4}\pin{k}\pin{p}\frac{(\omega_p-\omega_k)^2}{\omega_p}\frac{\tilde{g}^2_{k}(p)}{C_{k}^2}
-\frac{1}{4}\pin{p}\frac{\tilde{g}_{B}(p)\tilde{g}_{B}(p)}{C_{B}^2}\omega_p\nonumber
\eea
is a scalar.

The Hamiltonian is seen to be just a sum of quantum harmonic oscillators described by $b$ and $b^\dag$ plus a center of mass motion described by $\phi_0$ and $\pi_0$.   The lowest energy state $\co|0\rangle$ therefore is the unique state which satisfies
\beq
b_k\co|0\rangle_0=\pi_0\co|0\rangle_0=0 \label{coeq}
\eeq
and it has energy $E_K=E_{cl}+Q$ by (\ref{quick}) and (\ref{clpt}) because
\beq
H\p\co|0\rangle_1=(E_{cl}+H_{PT})\co|0\rangle_0=(E_{cl}+Q)\co|0\rangle_0.
\eeq
The excited states are just the oscillator excitations, made from products of $b^\dag_k$, and arbitrary momenta may be considered within the validity of the one-loop approximation.

Numerically evaluating $Q$, we find
\beq
Q_C=-0.034091m \hsp
Q_{B}=-0.284219m\hsp
Q=-0.318310m\hsp
\eeq
which agrees with the result $Q=-m/\pi$ obtained in Ref.~\cite{luther} using, essentially, the integrability \cite{johnson73,ft} of the Sine-Gordon model.

\section{Conclusion}


We used the Sine-Gordon model to test the method introduced in Ref.~\cite{mekink} for the calculation of the one-loop correction to soliton masses.  While the WKB method has been applied to both models \cite{dhn2,dhnsg} it suffers from an ambiguity due to a choice of matching of regularization conditions \cite{re}.  However in the case of the Sine-Gordon model, the soliton mass has been calculated unambiguously using integrability in Ref.~\cite{luther}.  Therefore, the case treated in this paper provides a robust test of our method.

The quantum soliton in the Sine-Gordon model is also of intrinsic interest.  As the Sine-Gordon model is understood at strong coupling, where it becomes the massive Thirring model \cite{colemansg}, it may be possible to follow the soliton operator to strong coupling. At one loop the operator may be found by solving (\ref{coeq}) for $\co$.   Of course it is well-known that in the Thirring model it becomes the fundamental fermion \cite{mandelop}, but it would be interesting to see what it becomes in terms of the strongly coupled Sine-Gordon model itself.  Perhaps this would give a hint as to what becomes of $\mathcal{N}=2$ SQCD monopoles \cite{sw2} when the Higgs mass tends to zero and so the scalar condensate turns off and the infrared coupling becomes strong?

\section* {Acknowledgement}

\noindent
JE is supported by the CAS Key Research Program of Frontier Sciences grant QYZDY-SSW-SLH006 and the NSFC MianShang grants 11875296 and 11675223.   JE also thanks the Recruitment Program of High-end Foreign Experts for support.


\begin{thebibliography}{99}

\bibitem{dhn2}
  R.~F.~Dashen, B.~Hasslacher and A.~Neveu,
  ``Nonperturbative Methods and Extended Hadron Models in Field Theory 2. Two-Dimensional Models and Extended Hadrons,''
  Phys.\ Rev.\ D {\bf 10} (1974) 4130.
  doi:10.1103/PhysRevD.10.4130
  
\bibitem{dhnsg}
  R.~F.~Dashen, B.~Hasslacher and A.~Neveu,
  ``The Particle Spectrum in Model Field Theories from Semiclassical Functional Integral Techniques,''
  Phys.\ Rev.\ D {\bf 11} (1975) 3424.
  doi:10.1103/PhysRevD.11.3424


\bibitem{colemansg}
  S.~R.~Coleman,
  ``The Quantum Sine-Gordon Equation as the Massive Thirring Model,''
  Phys.\ Rev.\ D {\bf 11} (1975) 2088.
  doi:10.1103/PhysRevD.11.2088

\bibitem{luther}
  A.~Luther,
  ``Eigenvalue spectrum of interacting massive fermions in one-dimension,''
  Phys.\ Rev.\ B {\bf 14} (1976) 2153.
  doi:10.1103/PhysRevB.14.2153


\bibitem{re}
  A.Rebhan and P.Van Nieuwenhuizen,
  ``No saturation of the quantum Bogomolnyi bound by two-dimensional supersymmetric solitons,''
  [hep-th/9707163]

\bibitem{mekink}
 J.~Evslin,
  ``Manifestly Finite Derivation of the Quantum Kink Mass,''
  JHEP {\bf 1911} (2019) 161
  doi:10.1007/JHEP11(2019)161
  [arXiv:1908.06710 [hep-th]].

\bibitem{hui}
  H.~Liu, Y.~Zhou and J.~Evslin,
  ``Ground States of the $\phi^4$ Double-Well QFT,''
  arXiv:1909.04946 [hep-th].

\bibitem{taylor78}
  J.~G.~Taylor,
  ``Solitons as Infinite Constituent Bound States,''
  Annals Phys.\  {\bf 115} (1978) 153.
  doi:10.1016/0003-4916(78)90179-3


\bibitem{hepp}
  K.~Hepp,
  ``The Classical Limit for Quantum Mechanical Correlation Functions,''
  Commun.\ Math.\ Phys.\  {\bf 35} (1974) 265.
  doi:10.1007/BF01646348






\bibitem{flugge}
S. Fl\"ugge,
``Practical Quantum Mechanics,"
Springer-Verlag Berlin Heidelberg (1999),
doi:10.1007/978-3-642-61995-3

\bibitem{johnson73}
  J.~D.~Johnson, S.~Krinsky and B.~M.~McCoy,
  ``Vertical-Arrow Correlation Length in the Eight-Vertex Model and the Low-Lying Excitations of the X-Y-Z Hamiltonian,''
  Phys.\ Rev.\ A {\bf 8} (1973) 2526.

\bibitem{ft}
  L.~D.~Faddeev, L.~A.~Takhtajan and V.~E.~Zakharov,
  ``Complete description of solutions of the Sine-Gordon equation,''
  Dokl.\ Akad.\ Nauk Ser.\ Fiz.\  {\bf 219} (1974) 1334
   [Sov.\ Phys.\ Dokl.\  {\bf 19} (1975) 824].


\bibitem{mandelop}
  S.~Mandelstam,
  ``Soliton Operators for the Quantized Sine-Gordon Equation,''
  Phys.\ Rev.\ D {\bf 11} (1975) 3026.
  doi:10.1103/PhysRevD.11.3026

\bibitem{sw2}
  N.~Seiberg and E.~Witten,
  ``Electric - magnetic duality, monopole condensation, and confinement in N=2 supersymmetric Yang-Mills theory,''
  Nucl.\ Phys.\ B {\bf 426} (1994) 19
   Erratum: [Nucl.\ Phys.\ B {\bf 430} (1994) 485]
  doi:10.1016/0550-3213(94)90124-4, 10.1016/0550-3213(94)00449-8
  [hep-th/9407087].


\end{thebibliography}
\end{document}